\documentclass[12pt]{article}   
\usepackage{epsfig}

\newlength{\dinwidth} \newlength{\dinmargin}
\begin{document}
\begin {flushright}
Cavendish-HEP-04/16\\
LBNL-55137
\end {flushright} 
\vspace{3mm}

\begin{center}
{\Large \bf Beauty and Charm Production in Fixed Target 
Experiments\footnote{Presented
at the DIS 2004 Workshop, Strbske Pleso, Slovakia, 14-18 April, 2004.}}
\end{center}
\vspace{2mm}

\begin{center}
{\large Nikolaos Kidonakis$^a$ and Ramona Vogt$^b$}\\
\vspace{2mm}
{\it $^a$Cavendish Laboratory, University of Cambridge\\
Madingley Road, Cambridge CB3 0HE, UK\\
E-mail: kidonaki@hep.phy.cam.ac.uk\\
\vspace{1mm}
$^b$Nuclear Science Division,\\
    Lawrence Berkeley National Laboratory, Berkeley, CA 94720, USA \\
and \\ Physics Department,
    University of California at Davis, Davis, CA 95616, USA\\
E-mail: vogt@lbl.gov}
\end{center}

\vspace{3mm}

\begin{abstract}

We present calculations of NNLO threshold corrections for 
beauty and charm production in $\pi^- p$ and $pp$ interactions
at fixed-target experiments.

\end{abstract}

\thispagestyle{empty} \newpage \setcounter{page}{2}

\section{Introduction} 

Recent calculations for heavy quark hadroproduction 
have included next-to-next-to-leading-order (NNLO) soft-gluon 
corrections \cite{NKtop} to the double differential cross section  
from threshold resummation techniques \cite{KS}.
These corrections are important for near-threshold 
beauty and charm production at fixed-target experiments, including HERA-B 
and some of the current and future heavy ion experiments.  

Soft-gluon corrections dominate the cross section near threshold.
They take the form of logarithms,
$[\ln^l(x_{\rm th})/x_{\rm th}]_+$, with $l \le 2n-1$ for the 
order $\alpha_s^n$ corrections, where $x_{\rm th}$ is a kinematical
variable that measures distance from threshold. 
In NNLO calculations ($n=2$) we denote leading logarithms 
(LL) with $l=3$, next-to-leading logarithms (NLL) with $l=2$, 
and next-to-next-to-leading logarithms (NNLL) with $l=1$.
The latest calculation used the methods of Ref. \cite{NKuni} 
to include next-to-next-to-next-to-leading
logarithms (NNNLL, $l=0$) at NNLO \cite{KVtop,KVbc}.
These NNNLL terms minimize the kinematics and scale dependence 
of the cross section.

Our calculation is done in single-particle-inclusive (1PI) 
kinematics since, in this kinematics, the NLO threshold approximation to the 
full NLO result is very good, as shown in Ref. \cite{KVbc}.
In 1PI kinematics, we
define $s=(p_a+p_b)^2$, $t_1=(p_b-p_1)^2-m^2$, $u_1=(p_a-p_1)^2-m^2$
and $s_4=s+t_1+u_1$ for the process 
$i(p_a) + j(p_b) \longrightarrow Q(p_1) + X[{\overline Q}](p_2)$
with $ij=q{\bar q}$ or $gg$.
At threshold $s_4 \rightarrow 0$ and the soft corrections take the form
$[(\ln^l(s_4/m^2))/s_4]_+$.

\section{Beauty production}

\begin{figure}[!thb]
\vspace*{10.0cm}
\begin{center}
\includegraphics{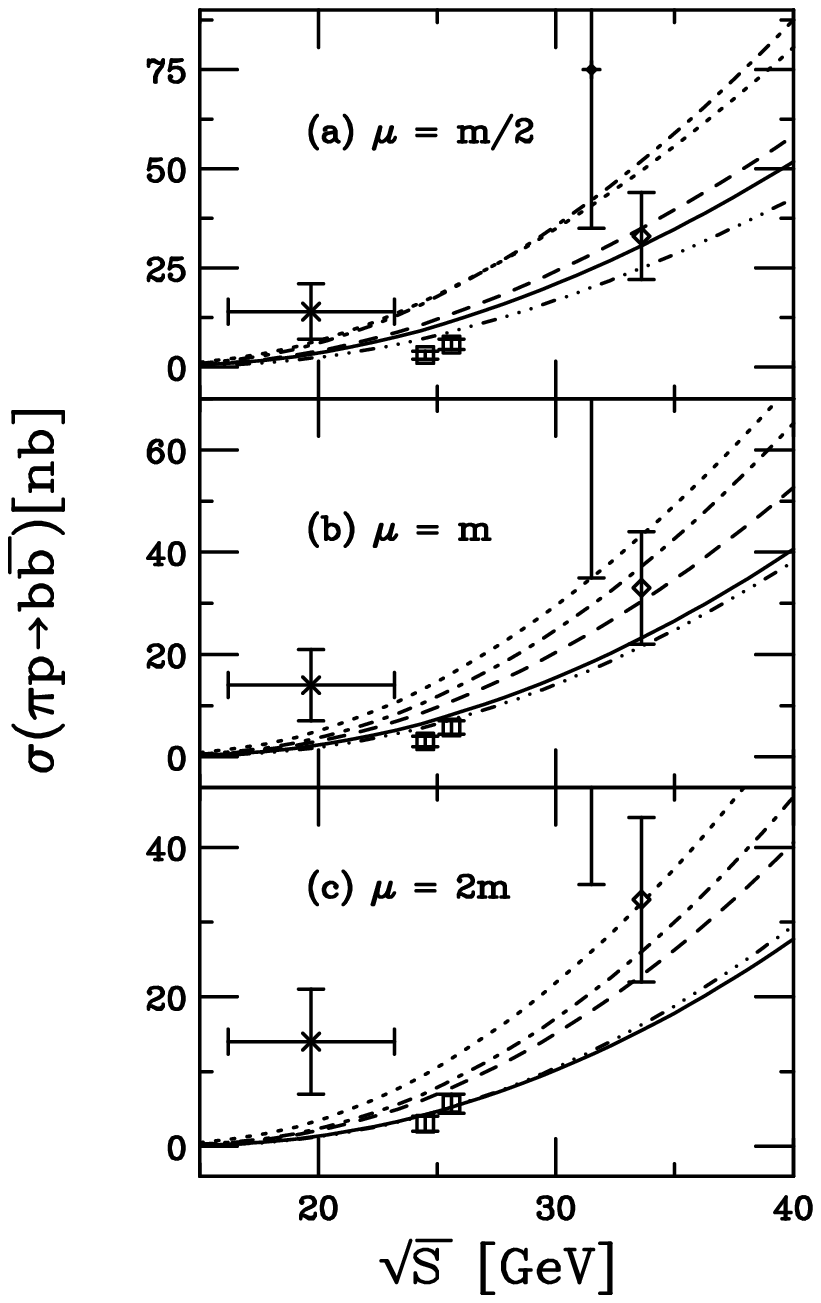}
\includegraphics{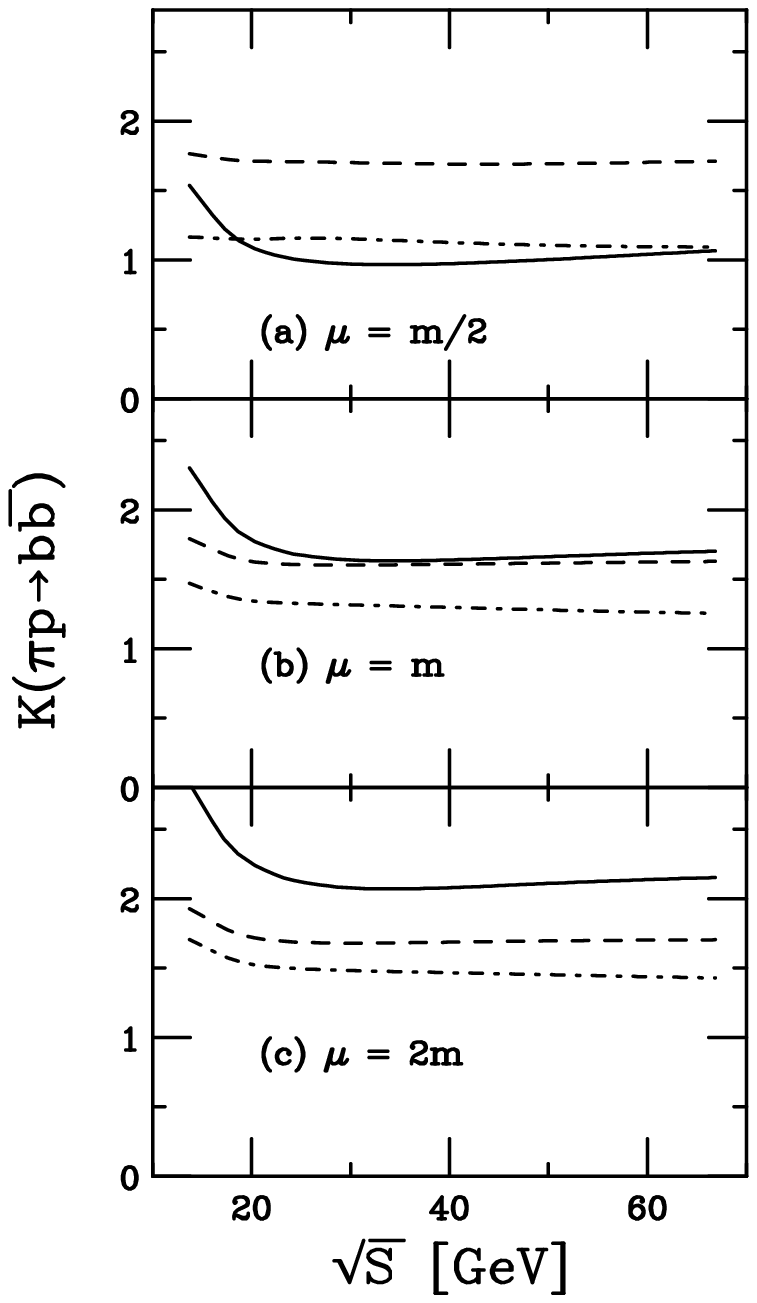}
\vspace{1.5cm}
\caption[*]{Beauty production cross sections and $K$ factors
in $\pi^- p$ interactions.}
\label{fig1}
\end{center}
\end{figure}

There is not much data on beauty hadroproduction at fixed-target
energies.  The $\pi^- p$ data \cite{na10,wa78,wa92,e653,e672} 
are shown on the left-hand side of Fig.~\ref{fig1} along with our calculations
with several choices of bottom quark mass and scale.  
We use the GRV98 HO proton parton densities \cite{grv98} 
with the GRS pion densities \cite{grs}.  For our central value of the 
bottom quark mass, $m = 4.75$ GeV, we present the exact NLO cross 
section (solid curve), 
the 1PI NNLO-NNLL cross section (dot-dashed) and the 1PI NNLO-NNNLL$+\zeta$ 
cross section (dashed).  
Here NNLO-NNLL indicates that we include the NNLL terms at NNLO, while,
for NNLO-NNNLL+$\zeta$, we include the NNNLL terms and some 
virtual $\zeta$ terms \cite{KVbc}.
We also show the 1PI NNLO-NNNLL$+\zeta$ cross sections
for $m = 4.5$ GeV (dotted) and 5 GeV (dot-dot-dot-dashed). 
On the right-hand side of Fig.~\ref{fig1} we present the $K$ factors 
for $m = 4.75$ GeV.
We show $\sigma_{\rm NLO}/\sigma_{\rm LO}$ (solid), 
$\sigma_{\rm NNLO-NNLL}/\sigma_{\rm NLO}$ (dashed), and 
$\sigma_{{\rm NNLO-NNNLL}+\zeta}/\sigma_{\rm NLO}$ (dot-dashed).

\begin{figure}[!thb]
\vspace*{10.0cm}
\begin{center}
\includegraphics{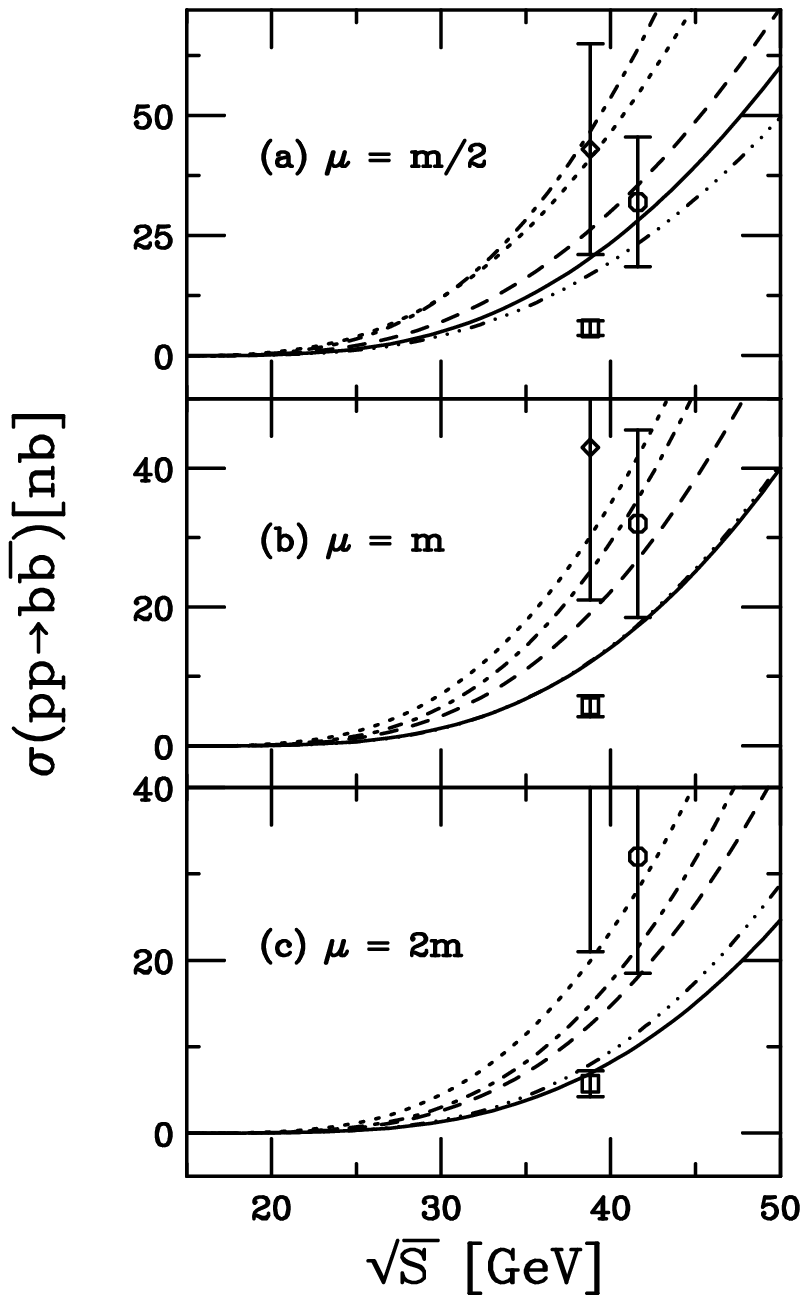}
\includegraphics{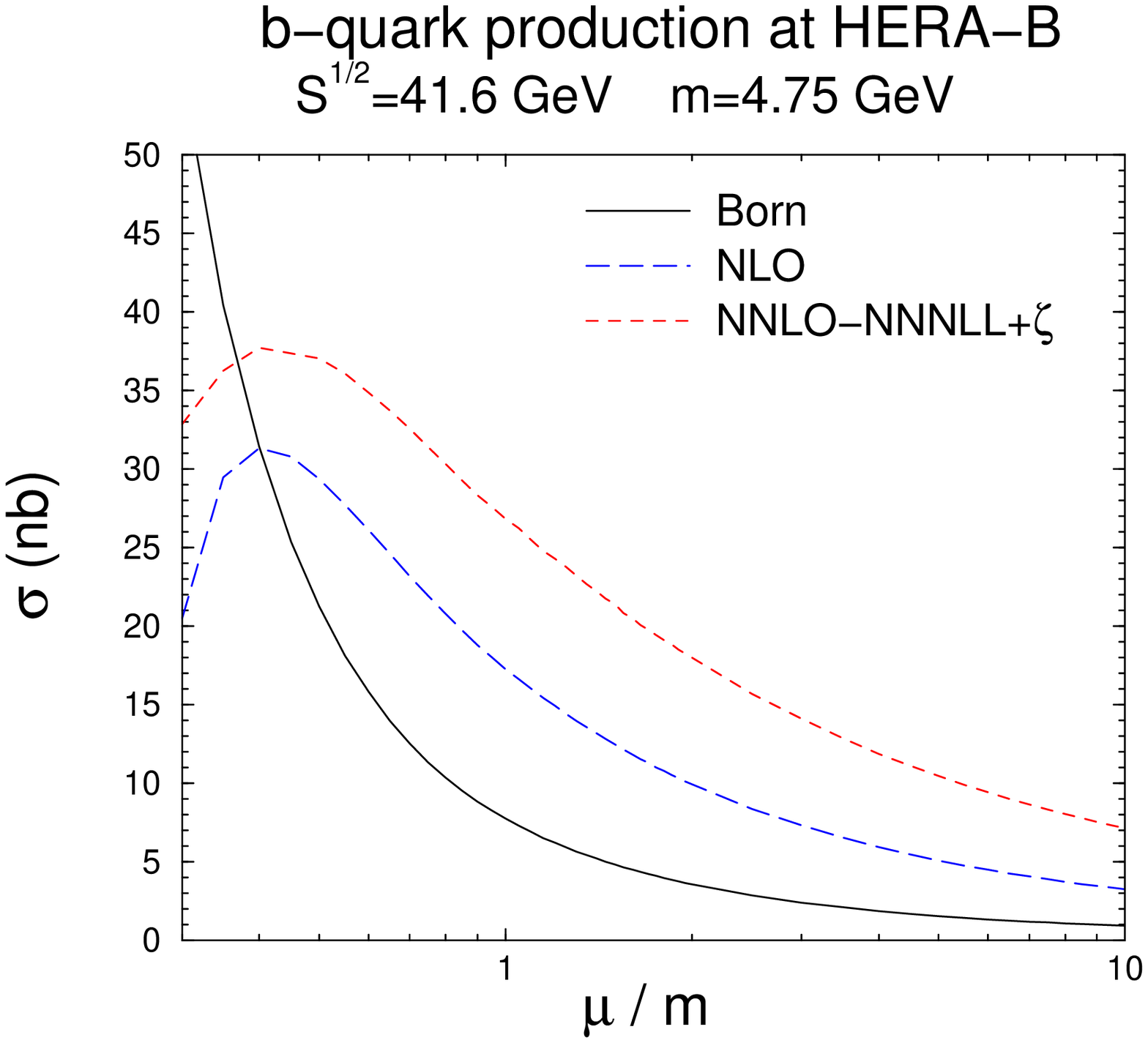}
\vspace{1.5cm}
\caption[*]{Beauty production in (left) $pp$ interactions
and (right) at HERA-B.}
\label{fig2}
\end{center}
\end{figure}

We now turn to beauty production in $pp$ interactions.
The data points from three experiments \cite{e789,e771,herab1}
are compared to our calculations 
with the MRST2002 NNLO parton densities \cite{mrst2002} 
on the left-hand side of Fig.~\ref{fig2}.

On the right-hand side of Fig.~\ref{fig2}, we plot the scale ($\mu$) 
dependence of the cross section at HERA-B for $0.3 < \mu/m < 10$ 
with $\sqrt{S}=41.6$ GeV and $m=4.75$ GeV. We show results for the 
Born, NLO, and NNLO-NNNLL+$\zeta$ cross sections.  
The scale dependence decreases with increasing order of the cross section.  
The plateau at $\mu/m \approx 0.4$ is broader for the 
NNLO-NNNLL$+\zeta$ cross section and the overall scale dependence is reduced
relative to the exact NLO cross section.

The NNLO-NNNLL$+\zeta$ $b \overline b$ cross section at $\sqrt{S} = 41.6$ GeV 
and $\mu=m=4.75$~GeV with the MRST2002 NNLO parton densities is
$\sigma_{{\rm NNLO-NNNLL}+\zeta}^{\rm MRST2002 NNLO} = 28 \pm 9
{}^{+15}_{-10}$~nb. 
The GRV98 densities give
$\sigma_{{\rm NNLO-NNNLL}+\zeta}^{\rm GRV98} = 25 {}^{+7}_{-8}
{}^{+13}_{-9}$ nb.  The first uncertainty is due to the scale variation,
$m/2 \le \mu \le 2m$, while the second is due to mass variation,
4.5 GeV $\le m \le$ 5 GeV.

Finally, we note that we find a reduction of the scale dependence for the 
NNLO-NNNLL$+\zeta$ $b{\bar b}$ cross section over all energies for 
both $\pi^- p$ and $pp$ interactions \cite{KVbc}.

\section{Charm production}

We now turn to charm quark production.  There is much more data on charm than
bottom production.
  
\begin{figure}[!thb]
\vspace*{10.0cm}
\begin{center}
\includegraphics{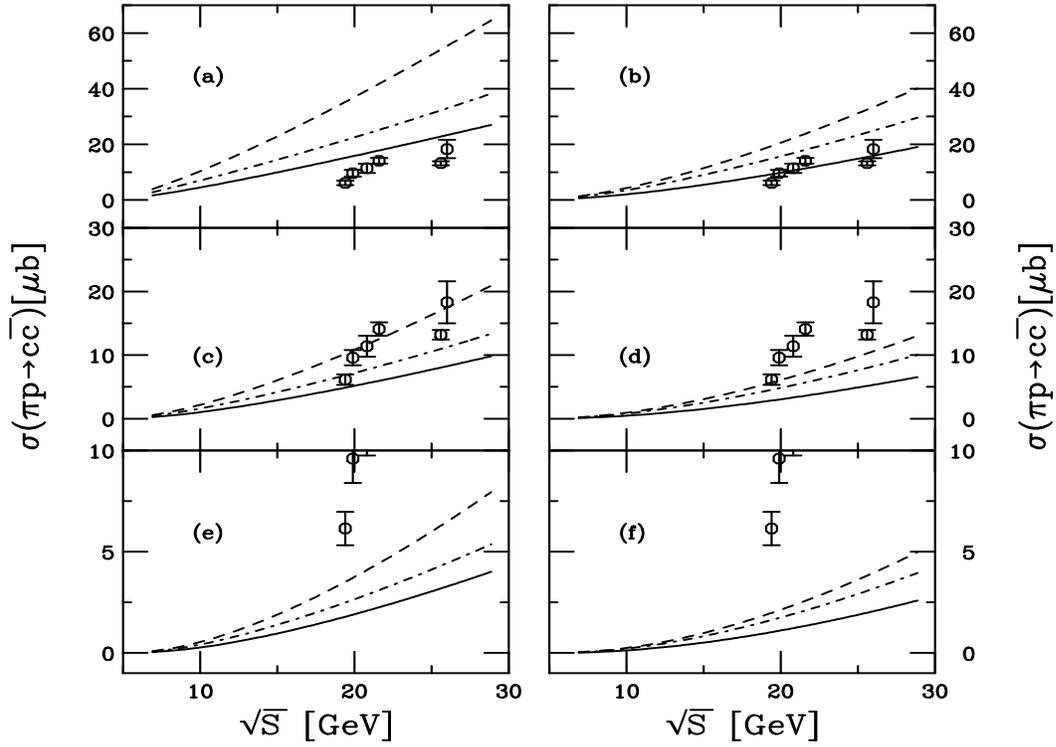}
\vspace{1.5cm}
\caption[*]{Charm quark production in $\pi^- p$ interactions.}
\label{fig3}
\end{center}
\end{figure}

In Fig.~\ref{fig3} we compare the $\pi^- p$ data from 
Refs.~\cite{Barlag88,Alves96,Barlag91,Adamovich96,Aguilar85a} with the 
exact NLO (solid), 1PI NNLO-NNLL (dashed) and 1PI
NNLO-NNNLL$+\zeta$ (dot-dashed) cross sections, calculated with the 
GRV98 HO proton parton densities and the GRS pion parton densities.
The mass of the charm quark is 1.2 GeV in (a) and (b), 1.5 GeV in (c) and (d),
and 1.8 GeV in (e) and (f). On the left-hand side $\mu=m$ while, on the
right, $\mu=2m$.

\begin{figure}[!thb]
\vspace*{10.0cm}
\begin{center}
\includegraphics{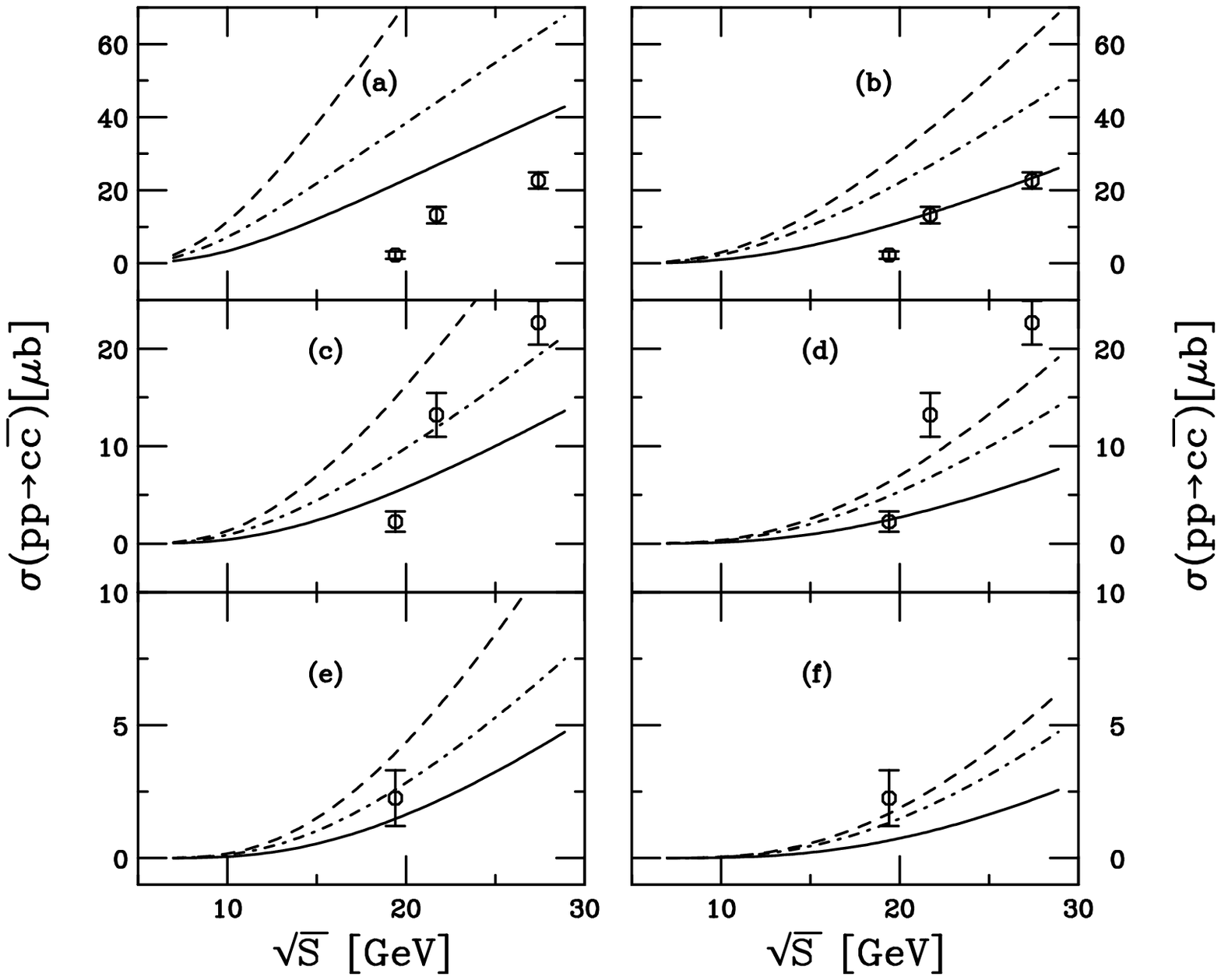}
\vspace{1.5cm}
\caption[*]{Charm quark production in $pp$ interactions.}
\label{fig4}
\end{center}
\end{figure}

Finally, we consider $pp \rightarrow c \overline c$ interactions.  In
Fig.~\ref{fig4}, we compare the data from 
Refs.~\cite{Barlag88,Alves96,Aguilar88} with 
exact NLO, 1PI NNLO-NNLL and 1PI
NNLO-NNNLL$+\zeta$ cross sections calculated with the MRST2002
NNLO proton parton densities.

We note that the $K$ factors are larger for charm than for 
beauty production and that the reduction of the scale dependence
is not as large \cite{KVbc}.

\section*{Acknowledgements} 

The research of N.K. has been 
supported by a Marie Curie Fellowship of the European Community programme 
``Improving Human Research Potential'' under contract no.
HPMF-CT-2001-01221. 
The research of R.V. is supported in part by the 
Division of Nuclear Physics of the Office of High Energy and Nuclear Physics
of the U.S. Department of Energy under Contract No. DE-AC-03-76SF00098.

\end{document}